# cDVAE: Multimodal Generative Conditional Diffusion Guided by Variational Autoencoder Latent Embedding for Virtual 6D Phase Space Diagnostics

A. Scheinker[*]

*Los Alamos National Laboratory, Los Alamos, New Mexico 87544, USA, [*]ascheink@lanl.gov*

## Abstract

Imaging the 6D phase space of a beam in a particle accelerator in a single shot is currently impossible. Single shot beam measurements only exist for certain 2D beam projections and these methods are destructive. A virtual diagnostic that can generate an accurate prediction of a beam's 6D phase space would be incredibly useful for precisely controlling the beam. In this work, a generative conditional diffusion-based approach to creating a virtual diagnostic of all 15 unique 2D projections of a beam's 6D phase space is developed. The diffusion process is guided by a combination of scalar parameters and images that are converted to low-dimensional latent vector representation by a variational autoencoder (VAE). We demonstrate that conditional diffusion guided by a VAE (cDVAE) can accurately reconstruct all 15 of the unique 2D projections of a charged particle beam's 6D phase space for the HiRES compact accelerator.

## Introduction

Particle accelerators are complex scientific instruments used for various high energy physics, biology, chemistry, and material science experiments. For example, free electron lasers (FEL) can generate intense pulses of X-rays which are only tens of femtoseconds (fs) in duration, allowing for the imaging of dynamic processes. Novel advanced schemes are being designed which require more precise control over fine details of the accelerated beam than what has previously been achieved. At advanced light sources such as the Linac Coherent Light Source (LCLS) upgrade LCLS-II FEL, there is a goal of precisely controlling the energies of two electron bunches separated by only tens of femtoseconds to generate x-rays with precisely tuned energy differences in two color mode operation [1]. A two-color mode is also being developed at the Swiss-FEL where advanced nonlinear compression techniques have enabled the production of attosecond hard x-ray FEL pulses [2]. The European XFEL is utilizing superconducting radio frequency (RF) cavities and low-level RF controls to accelerate bunches at MHz repetition rates [3]. The plasma wakefield acceleration facility for advanced accelerator experimental tests (FACET) upgrade, FACET-II, is planning on providing custom tailored current profiles with fs-long bunches and hundreds to thousands of kA peak currents [4]. At CERN's Advanced Proton Driven Plasma Wake-field Acceleration Experiment (AWAKE) facility the goal is to use transversely focused high intensity ($3\times10^{11}$), high energy (400 GeV) protons from CERN's Super Proton Synchrotron (SPS) accelerator to create a 10-meter-long plasma and wake-fields into which 18.8 MeV 650 pC electron bunches will be injected and accelerated up to 2 GeV [5].

Precise control of intense charged particle beams in accelerators is challenging because the beams change with time, and accurate non-invasive beam measure are not available. Accelerators are composed of thousands of coupled complex electromagnetic devices used to accelerate and focus the beams. The input-output characteristics of these devices are time-varying due to environmental disturbances such as vibrations and temperature variations and due to misalignments, aging of equipment, and equipment-beam interactions. Furthermore, the beams themselves have time-varying initial conditions due to the time-varying nature of complex plasma-based beam sources or sources based on cathode-laser interactions in which both the characteristics of the cathodes and of the laser systems change with time. These difficulties are compounded by an inability to quickly and non-destructively measure the 6D ($x,y,z,p_x,p_y,p_z$) phase space of the beams, which is evolving in a complex way due to collective effects such as coherent synchrotron radiation and space charge forces which distort the phase space in a nonlinear way [6,7]. Recently, a novel approach demonstrated the world's first 6D beam measurement at the Spallation Neutron

Source Beam Test Facility revealing previously unknown correlations in the six-dimensional phase space distribution, however this is not a single-shot measurement, it relies on ~18 hours of repeated beam measurements [8].

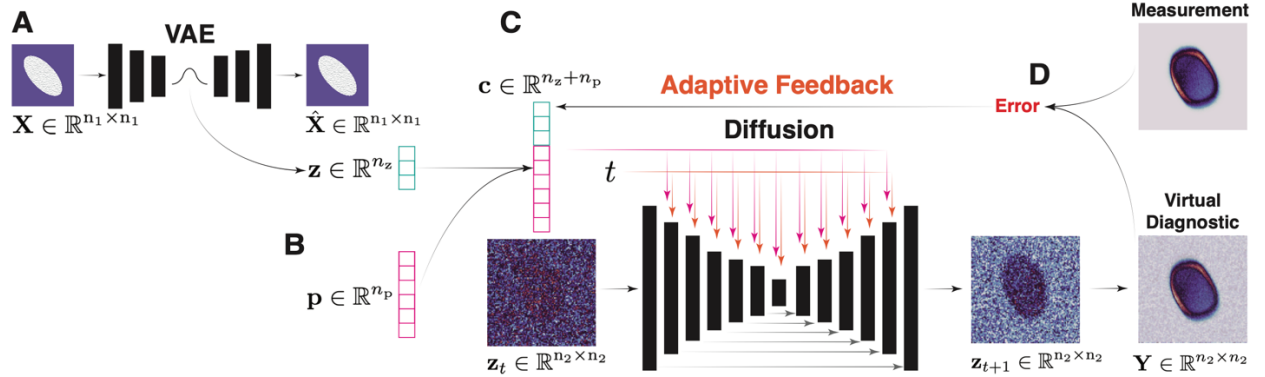

**Figure 1:** *A variational autoencoder (VAE) maps a $n_1 \times n_1$ image $X$ down into a low-dimensional ($n_z \ll n_1^2$) probabilistic latent embedding $z$, from which the image is then reconstructed (A). The latent embedding is concatenated together with a vector of accelerator and beam parameters $p$ (B) and the resulting vector c is used as a conditional guide for the diffusion model which generates high resolution ($n_2 \gg n_1$) images of the 15 unique projections of the beam's 6D phase space (C). If any of the generated projections are also available as online measurements, the difference between measured and generated distributions can be used to adaptively tune the conditional latent embedding $c$ to track the time-varying beam (D).*

*Summary of Main Results*

In this paper, a generative multimodal AI-based approach is presented for non-invasively generating all 15 unique 2D projections ((x, y), (x, z), … , (z, E)) of a beam's 6D phase space based on a conditionally guided diffusion process. In many accelerator applications the available inputs to a model or to a virtual diagnostic contains vectors of accelerator and beam parameters including RF cavity amplitude and phase set points, magnet power supply currents, current monitors, and beam position monitors. Sometimes there are also images that can be used as inputs, such as beam profile monitors at certain accelerator locations or TCAV-based longitudinal phase space (z, E) measurements. In this work, a multi-modal conditional diffusion guided by a VAE (cDVAE) method is developed that can combine images together with vectors to be used to guide the generative diffusion process, which can then generate incredibly high-resolution images as virtual beam diagnostics. The approach first embeds an image-based measurement $X$ which is typically a very high dimensional object into a low-dimensional latent space $z$. For example, even the relatively low resolution 52x52 pixel images of the beam's initial (x,y) distribution which are used in this work are $52^2$=2704-dimensional objects, which a VAE embeds into a 3-dimensional latent embedding. The latent embedding $z$ is then concatenated together with a vector $p$ of parameters which describe the accelerator and the beam into a single conditional vector $c$. The vector $c$ is then used to guide the diffusion process to generate very high resolution (256x256 pixels in this paper) images of projections of the beam's 6D phase space. The overall setup is shown in Figure 1. Besides being a natural way to combine different modalities of data into a common representation, this approach also enables unsupervised adaptive tuning of the generated phase space images by adaptive tuning of the latent conditional vector $c$, so that time-varying beams can be tracked over time even without access to the time-varying beam and accelerator inputs in $X$ and $p$ as was first demonstrated in [9]. Such latent space tuning also helps the generative model to extrapolate further beyond the span of the training data set than what is possible with traditional non-adaptive generative methods as was first demonstrated in [10].

## Background on Generative ML and Accelerator Beam Phase Space Diagnostics

The development of virtual and non-invasive phase space diagnostics has been a longstanding goal of the accelerator community. One of the first works focused on reconstructing the (z, E) longitudinal phase space (LPS) distribution of time-varying electron beams in the FACET beam-driven plasma wakefield accelerator combined an online physics model together with non-invasive energy spread spectrum measurements to track the measurements that otherwise required a destructive TCAV-based measurement [11]. The initial success of virtual diagnostics in [11] led to a series of further works on predicting a beam's 2D LPS including various neural network-based approaches at SLAC [12,13], and a novel approach that was developed at the EuXFEL in which a megapixel-resolution LPS diagnostic was developed for a wide range of beam conditions [14]. A novel invers-modeling method was then developed which combined generative ML together with an online physics model and adaptive feedback in order to map downstream beam measurements back to the initial (x,y) beam distribution at the accelerator beam source as well as tracking time-varying accelerator parameters [15]. Virtual diagnostics have also been developed for time stamping ultrafast electron diffraction beams by utilizing transformers combined with multilinear regression [16].

For higher dimensional predictions, the first generative model-based virtual 6D phase diagnostic was developed by LANL in [9], in which following training the low-dimensional latent space of an autoencoder convolutional neural network was adaptively tuned in an un-supervised approach that could track a time-varying unknown beam's phase space based on limited available diagnostics. A systematic study then showed that this approach could extrapolate much further beyond that span of the training data then standard feed-forward based ML approaches [10]. Recently, a machine learning approach for 4D transverse phase space tomography has been developed using image compression and machine learning [17], a neural network-based approach has been developed for predicting the 4D transverse emittance of space charge dominated beams using an advanced phase advance scan technique [18], 4D and 5D phase-space tomography techniques have been developed [19], 5D tomographic phase-space reconstruction methods have been studied [20].

In terms of state-of-the-art generative ML techniques, recently the first demonstration of using normalizing flow-based methods for high-dimensional phase space tomography has been developed [21] and the first demonstration of a conditional diffusion-based method has also been developed for a high-resolution (1024x1024 pixels) megapixel virtual diagnostic of the 2D LPS of the EuXFEL electron beam [22]. The utility of virtual phase space diagnostics for enabling control of charged particle beam phase space was already made clear back in 2018 when the first adaptive ML method was demonstrated at the LCLS FEL for automatic fast control of the electron beams LPS by combining a novel form of model-independent adaptive feedback [23], with neural network-based regression which could map desired 2D LPS designs directly to accelerator parameters and then adaptively tune them to zoom in on the correct accelerator settings and track them as the beam and accelerator drifted with time [24].

A major benefit of all neural network-based approaches is that trained neural networks are differentiable objects allowing for various optimization, design, and tuning studies by taking derivatives of the generated beam phase space distributions with respect to any of the model inputs.

Diffusion-based models are currently the state-of-the-art generative AI method for creating high resolution images. The diffusion approach is of gradually transforming images into analytically tractable Gaussian distributions and then teaching a neural network to denoising those images is an approach inspired by statistical thermodynamics for modeling complex distributions [25]. This approach was then further developed for the generation of high-resolution images [26-29]. Diffusion-based generative models have now become the state-of-the-art for generating high resolution images, especially when the images have a wide variety. The generative ability of diffusion-based models has made them powerful tools for a wide range of scientific applications [30], such as conditional generation of hypothetical new families of superconductors [31], for brain imaging [32], for generating multi-pathological and multi-modal images and labels for brain MRI [33], for various bioengineering applications [34], for protein structure generation [35], and for modeling electron-proton collisions in high energy physics [36].

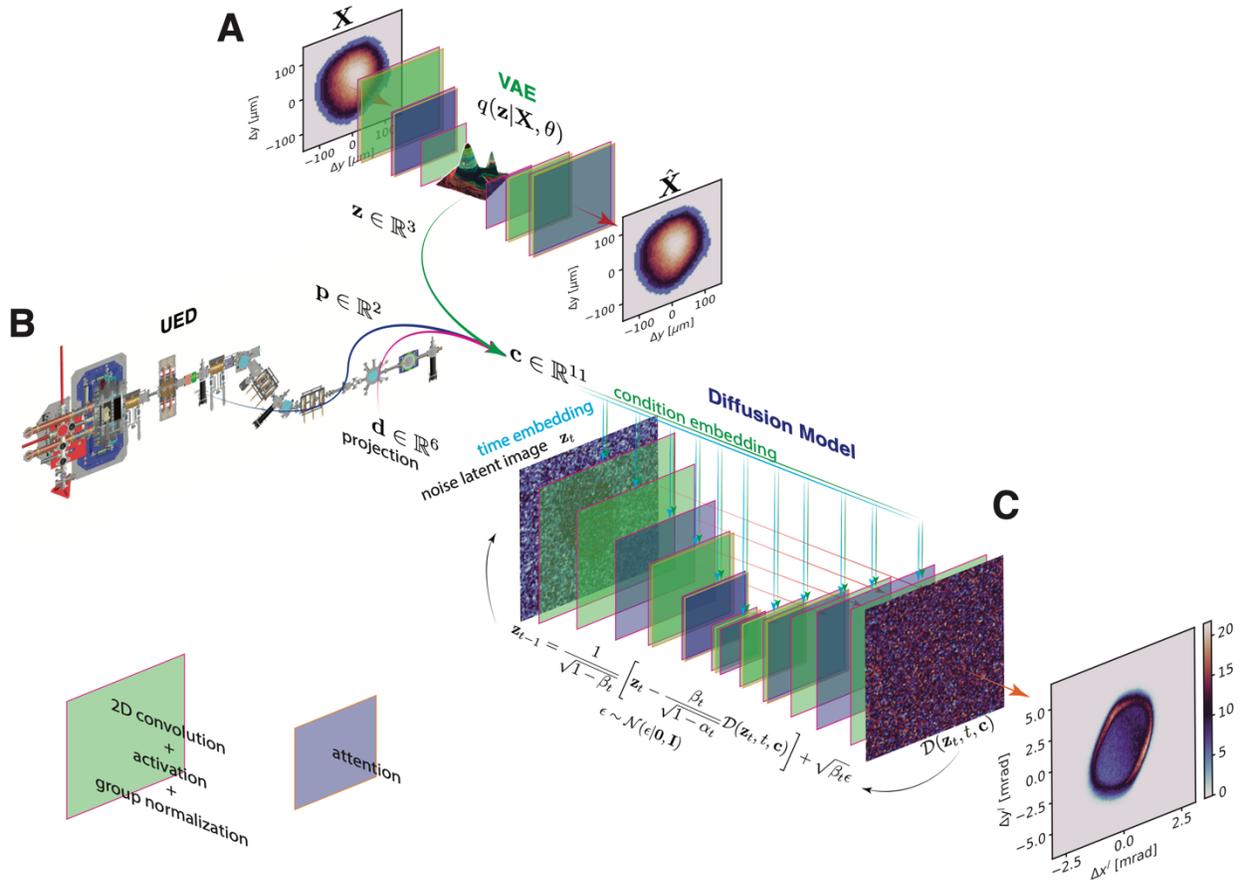

*Figure 2: The overall cDVAE setup is shown. **A:** A VAE is used to encode the 2D image of the beam's initial (x,y) distribution into a 3-dimensional latent space embedding **z**. **B:** The latent **z** is concatenated with a vector **p** of scalar parameters describing accelerator components and beam properties and a vector **d** which defines which of the 15 2D projections should be generated. **C:** An 11-dimensional conditional vector **c**=[z,p,d] is used to guide the diffusion process to generate individual 2D projections of the beam's 6D phase space.*

## cDVAE: Multimodal Conditional Diffusion Guided by VAE Latent Space

The diffusion model used in cDVAE is based on the denoising diffusion method [27], with the modification that a conditional vector is added with the time embedding to guide the diffusion process. Conditional diffusion is similar to the work in [22], with the major generalization of this work being the use of a VAE to also embed information about measurement images, rather than just vectors of beam and accelerator parameters as was done in [22], although that work generates much higher resolution images. Here it is worth noting that the idea of conditional guidance for generative models has also been explored in a recent novel approach of utilizing conditionally guided neural networks for errant beam prognostics [37].

The scalar parameters, $p$, are based on measurements such as magnet power supply current and bunch charge. The images $X$ are of experimentally measured (x,y) beam distributions at the accelerator entrance. To utilize these images in an efficient way a VAE first maps them into a low-dimensional latent vector representation $z$, and then the entire vector $\mathbf{c} = [p, z]$ is used to guide the generative diffusion process. The VAE learns a probabilistic representation $z$ of its inputs $X$ by using two neural network branches in the convolutional neural network-based encoder to learn mean and variance functions $f_\mu[X, \theta]$ and $f_\Sigma[X, \theta]$, which are combined to form a probabilistic density function over $z$ defined as the multivariate Gaussian:

$$q(\mathbf{z}|\mathbf{X}, \theta) = \mathcal{N}_z\left[f_\mu[\mathbf{X}, \theta], f_\Sigma[\mathbf{X}, \theta]\right].$$

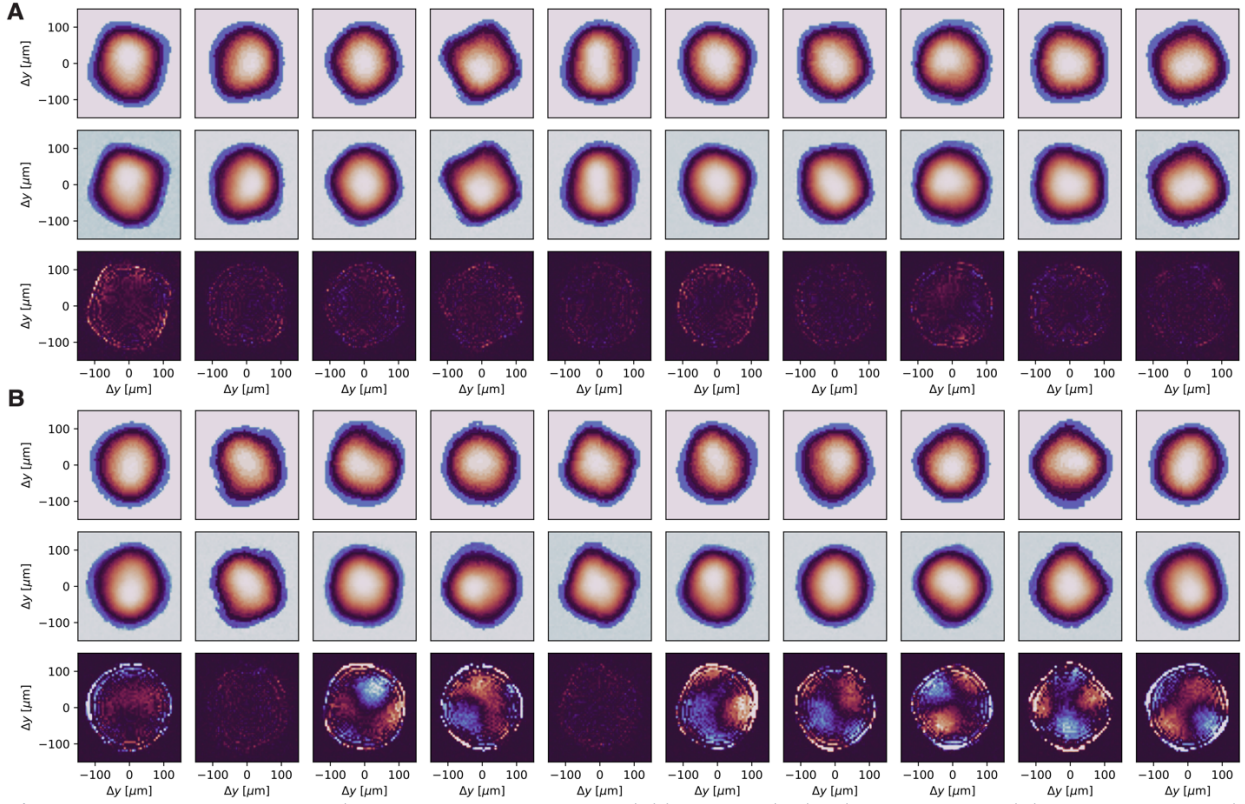

*Figure 3*: VAE-based reconstructions of 10 randomly chosen initial beam distributions from the training data are shown in (A) relative to 10 randomly chosen test data distributions in (B). The model is more accurate on the training data. On the test data it is able to capture all of the important features, missing only fine texture-like details.

In the above equation θ represents the learned weights of the VAE encoder. In this approach, an input image $X$ is first embedded into a mean and variance and then a random $z$ is sampled as a 3-dimensional vector from the multivariate distribution. The VAE decoder then regenerates the input image as a function of $z$:

$$\hat{X} = G[z, \theta],$$

where θ represents the weights of the decoder part of the VAE. The VAE training process samples random batches of input images $\beta = \{X_i, X_j, \ldots\}$, passes them as inputs to the VAE, and adjusts the VAE weights in order to maximize the log-likelihood or equivalently minimize the negative log-likelihood of the training data given the VAE weights. In addition to this reconstruction loss term the network also tries to minimize the Kullback-Leibler (KL) divergence between the latent distribution $p(z|X,\theta)$ and a specified distribution in the latent space [38]. The combination of the log-likelihood and KL divergence is known as evidence lower bound (ELBO), which by Jensen's inequality is strictly less than or equal to the log-likelihood. Considering the true unknown image generating distribution $Pr(\mathbf{X}|\varphi)$ parameterized by some unknown parameters $\varphi$, and the VAE-based distribution $q(z|X,\theta)$, the ELBO is given by

$$\text{ELBO}[\theta, \varphi] = \int q(\mathbf{z}|\mathbf{X}, \theta) \log \left[ Pr(\mathbf{X}|\mathbf{z}, \varphi) \right] d\mathbf{z} - D_{\text{KL}} \left[ q(\mathbf{z}|\mathbf{X}, \theta) || Pr(\mathbf{z}) \right]$$

The first term in the above is the log-likelihood of the training data given a choice of neural network weights $\theta$ and is an intractable high-dimensional integral that cannot be directly computed. This part of the ELBO is calculated by a Monte Carlo estimate-based sampling technique

$$\mathrm{E}_{q(\mathbf{z}|\mathbf{X}, \theta)}[\log[\Pr(\mathbf{X}|\mathbf{z}, \varphi)]] = \int q(\mathbf{z}|\mathbf{X}, \theta) \log[\Pr(\mathbf{X}|\mathbf{z}, \varphi)] \, d\mathbf{z} \approx \frac{1}{N} \sum_{n=1}^{N} \log[\Pr(\mathbf{X}|\mathbf{z}_n^*, \varphi)]$$

where the $z_n^*$ are samples from the learned distribution $q(z|X,\theta)$. Another major simplification is to compare the generated distribution $q(z|X,\theta)=N[\mathbf{f}_\mu, \mathbf{f}_\Sigma]$ to a simple normal distribution as the prior $Pr(z)=N_z[\mathbf{0}, \mathbf{I}]$. This simplifies the KL divergence to

$$\mathrm{D_{KL}}\left[q(\mathbf{z}|\mathbf{x},\theta)||Pr(\mathbf{z})\right] = \frac{1}{2}\left(\mathrm{Tr}[\mathbf{f}_\Sigma] + \mathbf{f}_\mu^T \mathbf{f}_\mu - D_\mathbf{z} - \log[\det[\Sigma]]\right)$$

where $D_z$ is the latent space dimension. The overall ELBO loss is then simplified into a computationally feasible calculation that can be carried out by sampling outputs of the VAE's latent embedding and generated images, based on the approximation

$$\mathrm{ELBO}[\theta,\varphi] \approx \log[Pr(X|z^*,\varphi)] - \frac{1}{2}\left(\mathrm{Tr}[\mathbf{f}_\Sigma] + \mathbf{f}_\mu^T \mathbf{f}_\mu - D_z - \log[\det[\mathbf{\Sigma}]]\right).$$

The overall setup is shown in Figure 2.

The first step of this approach was the choice of latent dimension size for the VAE-based image embedding. While a higher dimensional latent space makes it much easier for the VAE to accurately reconstruct the input images, a lower-dimensional latent space enables much faster adaptive iteration when tracking unknown time-varying initial beam conditions in the unsupervised application mode of the approach and lower dimensional latent embeddings are also easier to interpret. Previous studies of latent space dimension vs reconstruction accuracy have shown that for the HiRES UED initial condition beam (x,y) data, increasing the latent dimension beyond 2D quickly shows diminishing results [10]. Therefore, for this work a $n_z$=3D VAE latent space was chosen to slightly increase reconstruction accuracy while maintaining easy latent space visualization and interpretability. The VAE was then trained on a mixture of experimentally measured initial beam (x,y) distributions from the HiRES UED as well as synthetically generated images which were constructed based on random combinations of the PCA components learned from the experimentally measured beams and their rotations, as previously discussed in [15]. A demonstration of the VAE's capabilities on 10 random training and 10 random test data images is shown in Figure 3. As expected, the model performs better on the training data, but captures all the important features in the test data as well with most of the error coming from fine texture-like details.

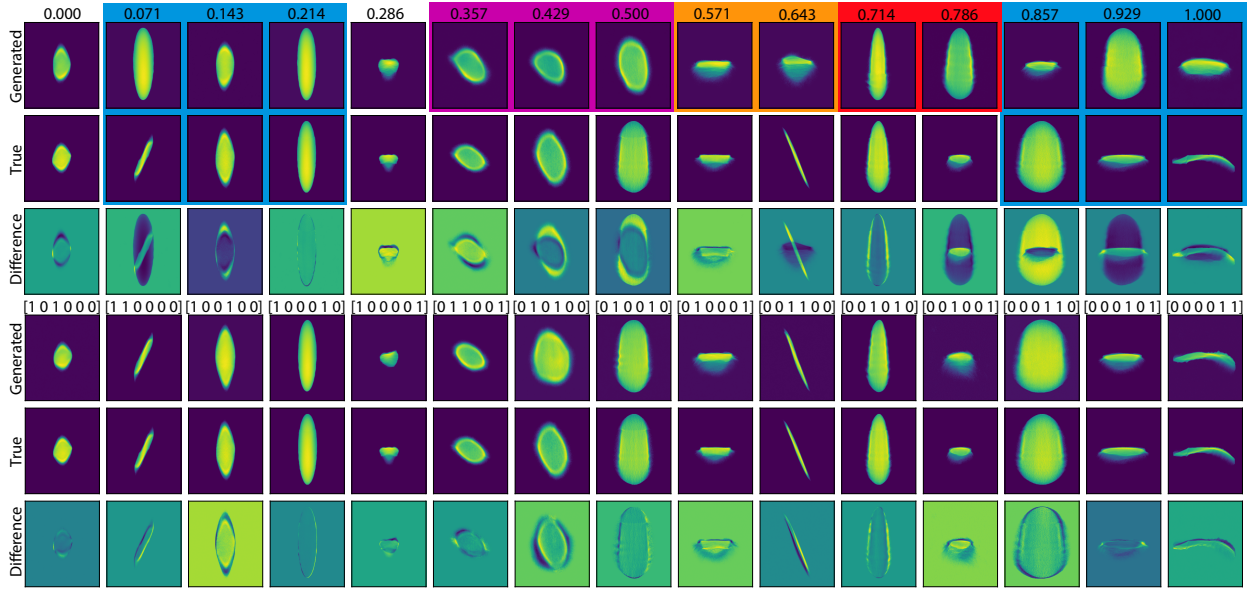

*Figure 4:* The top 3x15 grid shows generated and true phase space projections and their difference when conditioning which of the 15 projections should be generated by 15 numbers evenly distributed between 0 and 1. The bottom 3x15 grid shows the same information when conditioning with orthogonal physics-informed vectors.

The next important choice is to decide how to condition the diffusion-based model for generating the virtual predictions of the beam's 6D phase space via 2D projections downstream in the HiRES accelerator. Naively the most simple approach might be to define the conditional vector *c* as simply *c*=[*z*, *p*], a combination of the image embedding *z* and the accelerator/beam scalars embedding *p*. In this case the diffusion model would then generate all 15 projections of the beam's phase space as a single 256×256 pixel image with 15 channels, resulting in a 256×256×15 tensor, which is a very large object. This approach was found to be very slow to converge as the various channels of the generated object blead over into each other and make it more difficult for the diffusion UNet to disentangle the various phase space projections

The model was found to work much better if an additional conditional vector was introduced, *d*, which defined which of the 15 projections the diffusion model should generate, so that the diffusion model at any time only has to generate single channel 256×256×1 tensors based on a conditional input which encodes which of the beam's 15 2D projections should be generated according to *c*=[*z*, *p*, *d*]. Another crucial choice in this process is how to represent each of the 15 dimensions in the vector *d*. The simplest choice, and one which adds least overhead to the overall cDVAE UNet architecture is to use a one-dimensional scalar value of *d*. This was tested with 15 values of *d* uniformly spread within the range [0,1]. This simple choice immediately introduces a problem that is well known in the ML community when working with classification. By choosing *d* as simple numerical values that are evenly distributed, it arbitrarily places certain projections closer to each other. Considering the order that was chosen for this data, while the first projection (x,y) should somehow be "close" to the second projection (x,x'), as only 1 of the 2 dimensions has changed, there is no reason that (x,y) should be closer to (x,x') than (x,z) or (x,y'), or (x,E). When working with classification problems this is exactly why, for example, given 3 different classes, rather than encoding them as evenly distributed numbers [0, 1/2, 1], they are "one-hot" encoded so that the first class is represented as [1, 0, 0], the second as [0, 1, 0], and the third as [0, 0, 1]. That way the classes are represented as orthogonal vectors with equal distance between any pair. However, in this work, it is not smart to one-hot encode to such a set of 15-dimensional vectors, because we do want to preserve the fact that a 2D projection of (x, z) should somehow be closer to the projection (x, y) with which it shares one axis, than to the projection (y', E), where both axes are different. Therefore, the vector *d* was chosen in 6D as shown in the labels of the bottom part of Figure 4. We consider the ordering of axes to be {0,1,2,3,4,5} for {x,y,x',y',z,E}, and then we generate vectors who's entries are zero except for ones at the locations of the axes that are being projected. Several examples of the projection-vector pairs are then given by:

| (x,y), *d* = [110000] | (x,x'), *d* = [101000] | (x,y'), *d* = [100100] | (x,z), *d* = [100010] | (x,E), *d* = [100001] |
|---|---|---|---|---|
| (y,x'), *d* = [011000] | (y,y'), *d* = [010100] | (y,z), *d* = [010010] | (y,E), *d* = [010001] | (x',y'), *d* = [001100] |
| (x',z), *d* = [000110] | (x',E), *d* = [001001] | (y',z), *d* = [000101] | (y',E), *d* = [000101] | (z,E), *d* = [000011] |

This physics-informed orthogonalization of the various projections resulted in much higher accuracy of vDVAE-based image generation. To demonstrate this we trained the diffusion model with both the scalar and the 6D *d*-vector conditioning methods and the results for one test object are shown in Figure 4. While all of the projections in the vectorized conditioning approach are well separated, in the scalar conditioning approach the colored boxes highlight regions where the diffusion model is confused and generating incorrect projections based on proximity of inputs because of the natural tendency of the neural network to simply interpolate between similar values of *d* when the orthogonal property is not built in.

The generative diffusion process works by gradually adding noise to an image *Y* over many steps T until the pixel values of the image are indistinguishable from zero mean unit variance Gaussian noise, as shown when moving from right to left along any one of the rows of Figure 5. The way that data is transformed into noise is to start with an image *Y* and iteratively noise it according to steps

$$\mathbf{z}_1 = \sqrt{1-\beta_1}\mathbf{Y} + \sqrt{\beta_1}\epsilon_1, \quad \mathbf{z}_2 = \sqrt{1-\beta_2}\mathbf{z}_1 + \sqrt{\beta_2}\epsilon_2$$

$$\vdots$$

$$\mathbf{z}_t = \sqrt{1-\beta_t}\mathbf{z}_{t-1} + \sqrt{\beta_t}\epsilon_t, \quad \epsilon_t \sim \mathcal{N}(\mathbf{0},\mathbf{I}), \quad \beta_t \in [0,1]$$

In the above formula, β$_t$ is the noise schedule which controls the rate at which noise is added to the image. This diffusion process can be written in terms of probability densities given by

$$q(\mathbf{z}_1|\mathbf{x}) = \mathcal{N}_{\mathbf{z}_1}\left[\sqrt{1-\beta_1}\mathbf{Y}, \beta_1\mathbf{I}\right], \quad q(\mathbf{z}_t|\mathbf{z}_{t-1}) = \mathcal{N}_{\mathbf{z}_t}\left[\sqrt{1-\beta_t}\mathbf{z}_{t-1}, \beta_t\mathbf{I}\right]$$

where each noise step $z_t$ depends only on the previous stable state $z_{t-1}$, forming a Markov chain.

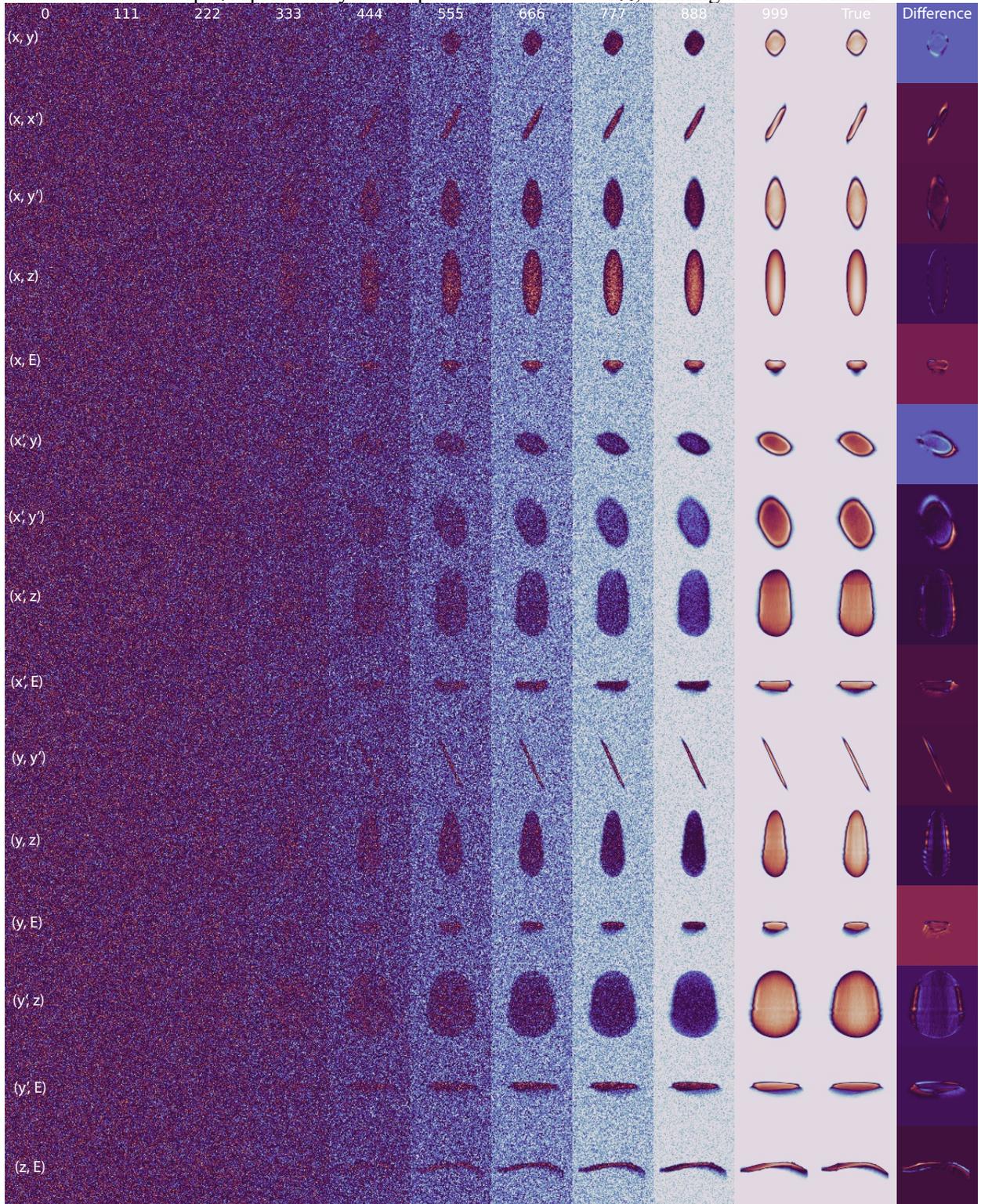

*Figure 5:* Every 111th step of the conditional diffusion process is shown generating all 15 projections of a beam's 6D phase space. The generated images are then compared to the true images and their difference is shown.

A joint distribution for all the sample states given a particular input $Y$ is then given by

$$q(z_{1...T}|Y) = q(z_1|Y) \prod_{t=2}^{T} q(z_t|z_{t-1}). \tag{1}$$

The diffusion model is trained to undo the noising process given batches of various samples $z_t$ for a given $Y$. To generate such samples during training, rather than running through the diffusion process sequentially, it is possible to derive the diffusion kernel $q(z_t|Y)$ of the process by calculating that for a given $Y$, repeated diffusion steps result in $z_t$ values of the form

$$z_t = \sqrt{\alpha_t}Y + \sqrt{1-\alpha_t}\epsilon, \quad \alpha_t = \prod_{s=1}^{t}(1-\beta_s) \implies q(z_t|Y) = \mathcal{N}_{z_t}[\sqrt{\alpha_t}Y, (1-\alpha_t)I]. \tag{2}$$

Theoretically, the diffusion kernel could then be used to calculate the marginal distribution $q(z_t)$ as

$$q(z_t) = \int q(z_t|Y)Pr(Y)dY, \tag{3}$$

but in practice this is impossible for a large space of images for which $Pr(Y)$ is unknown. In order to train the decoder, to sequentially undo the training process, for a given image $Y$ it is possible to calculate the conditional diffusion distribution $q(z_{t-1}|z_t,Y)$ by a combination of Baye's rule and a change of variables:

$$q(z_{t-1}|z_t, Y) = \mathcal{N}_{z_{t-1}}\left[\frac{1-\alpha_{t-1}}{1-\alpha_t}\sqrt{1-\beta_t}z_t + \frac{\sqrt{\alpha_{t-1}}\beta_t}{1-\alpha_t}Y, \frac{(1-\alpha_{t-1})\beta_t}{1-\alpha_t}I\right]. \tag{4}$$

The diffusion model is then constructed as a neural network that can probabilistically map $z_t$ back to $z_{t-1}$ iteratively until $Y$ is recovered. At this step a simplification is made in assuming that the various probability distributions involved can be approximated by normal distributions, such that

$$Pr(z_T) = \mathcal{N}_{z_T}[0, I],$$
$$Pr(z_{t-1}|z_t, \theta_t) = \mathcal{N}_{z_{t-1}}[D_t[z_t, \theta_t], \sigma_t^2 I],$$
$$Pr(Y|z_1, \theta_1) = \mathcal{N}_Y[D_t[z_1, \theta_1], \sigma_1^2 I], \tag{5}$$

where $D_t$ is a neural network with weights $\theta_t$ which is used to model the mean in an approach very similar to what was described above when using a VAE and unlike the VAE approach, the variance is pre-defined:

$$\sigma_t^2 = \frac{1-\alpha_{t-1}}{1-\alpha_t}\beta_t. \tag{6}$$

The neural network is itself a function of t which is passed into each layer, as shown in Figure 2, to condition the network on which time step it is currently taking. The loss function used to train the diffusion network is again based on the ELBO as described above for the VAE, given by

$$L = \sum_{j=1}^{N_b} \left( -\log\left[\mathcal{N}_{Y_j}[D_1[z_{j1}, \theta_1], \sigma_1^2 I]\right] \right.$$
$$\left. + \sum_{t=2}^{T} \frac{1}{2\sigma_t^2}\left\|\frac{1-\alpha_{t-1}}{1-\alpha_t}\sqrt{1-\beta_t}z_{jt} + \frac{\sqrt{\alpha_{t-1}}\beta_t}{1-\alpha_t}Y_j - D_t[z_{jt}, \theta_t]\right\|^2 \right), \tag{7}$$

where $N_b$ is the batch size, $Y_j$ is the jth image from the batch, and $z_{jt}$ are the latent variables associated with image $Y_j$ at diffusion step t as the network is trained in parallel over the batch. The first term inside of the batch sum in Equation (7) is the negative log-likelihood which is a metric of how accurate the reconstructed image is and the second term is comparing the neural network's prediction of the distribution's mean to the analytically predicted mean given in Equation (4). In practice, it has been found that instead of generating $z_{t-1}$ from $z_t$, as described above, the diffusion process has better performance if the network is instead used to try to predict the noise that was used to transform $z_{t-1}$ to $z_t$ and then that noise can be subtracted from $z_t$ to generate an estimate of $z_{t-1}$. In order to do this, we rearrange the diffusion process as

$$\mathbf{z}_t = \sqrt{\alpha_t}\mathbf{Y} + \sqrt{1-\alpha_t}\epsilon \quad \Longrightarrow \quad \mathbf{Y} = \frac{1}{\sqrt{\alpha_t}}\mathbf{z}_t - \frac{\sqrt{1-\alpha_t}}{\sqrt{\alpha_t}}\epsilon. \tag{8}$$

Plugging this representation of $Y$ into the mean term of Equation (7) and rearranging terms gives

$$L = \sum_{j=1}^{N_b} \left(-\log\left[\mathcal{N}_{\mathbf{Y}_j}[\mathbf{D}_1[\mathbf{z}_{j1},\theta_1],\sigma_1^2\mathbf{I}]\right]\right.$$
$$\left. + \sum_{t=2}^{T} \frac{1}{2\sigma_t^2}\left\|\left(\frac{1}{\sqrt{1-\beta_t}}\mathbf{z}_{jt} - \frac{\beta_t}{\sqrt{1-\alpha_t}\sqrt{1-\beta_t}}\epsilon\right) - \mathbf{D}_t[\mathbf{z}_{jt},\theta_t]\right\|^2\right). \tag{9}$$

Next, we modify the above approach by adding an additional conditional input to $\mathbf{D}_t$ which is the conditional vector $c$ containing projection choice, accelerator and beam parameters, and the latent image embedding, which is generated by the VAE, as described above. Finally, to use the model itself to predict the noise term, $\mathbf{D}_t$ is modified according to

$$\frac{1}{\sqrt{1-\beta_t}}\mathbf{z}_t - \frac{\beta_t}{\sqrt{1-\alpha_t}\sqrt{1-\beta_t}}\mathbf{D}_t[\mathbf{z}_t,\theta_t,\mathbf{c}=[\mathbf{p},\mathbf{z},\mathbf{d}]] \tag{10}$$

so that the job of the neural network is now simplified to that of predicting the noise at each step t.

With this training architecture for the cDVAE chosen, the model is then trained on 3000 training data sets in which each training object is a set of 15 projections of a beam's 6D phase space, each projection is a 256×256 pixel object. The input (x,y) distributions are chosen from the training and test set of the VAE and the beam and the accelerator parameters $p$ are varied over a grid, with the beam's charge varying with 10 solenoid currents ranging from 4.65 to 4.85 A and 4 bunch charges of 0.25, 0.5, 0.75, and 1 pC. One example of the cDVAE generative process, creating all 15 projections of a beam's 6D phase space for a test object is shown in Figure 5 alongside the correct projections. It is not surprising that the network is so accurate for a test object because in this case the problem is of interpolation over the range of the data set. A statistical summary of the test and training data is shown in Figure 6 in blue and yellow.

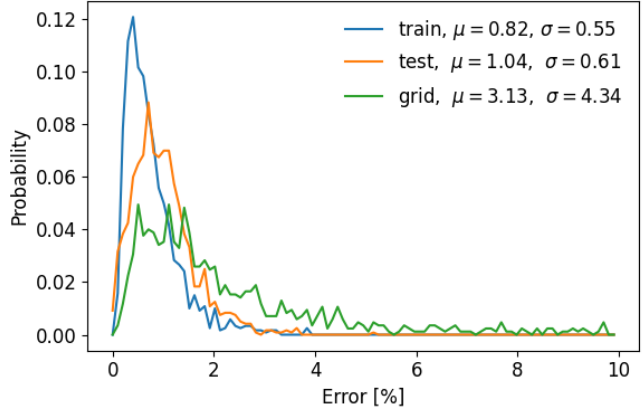

*Figure 6*: Percent absolute error averaged over all 15 projections for the test, training, and grid-based data.

Having confirmed that cDVAE can accurately interpolate, the much more interesting and more challenging problem is to test extrapolation at points beyond the grid. A wider range 6×10 grid was then set up, as shown in Figure 7. The distribution of percent absolute error averaged over all 15 projections for the wider grid data is shown in Figure 6 and the points are also colored by this error in Figure 7. The model does very well for grid points within the range of the training data and is also able to extrapolate somewhat, although it does break down at the far edges beyond what it has seen.

The grid is traversed starting point at the bottom left corner of Figure 7, which is then interpolated over 60 steps to the grid point at which a 90-degree left turn is made. The path then stops at every vertical point in that column with 60 steps between points, until making a 90 degree right turn and taking a final 60 steps to reach the top right. In total, along the path 8 grid points are visited to check the model's ability to make predictions of the 6D phase space. These predictions are summarized in Figure 8.

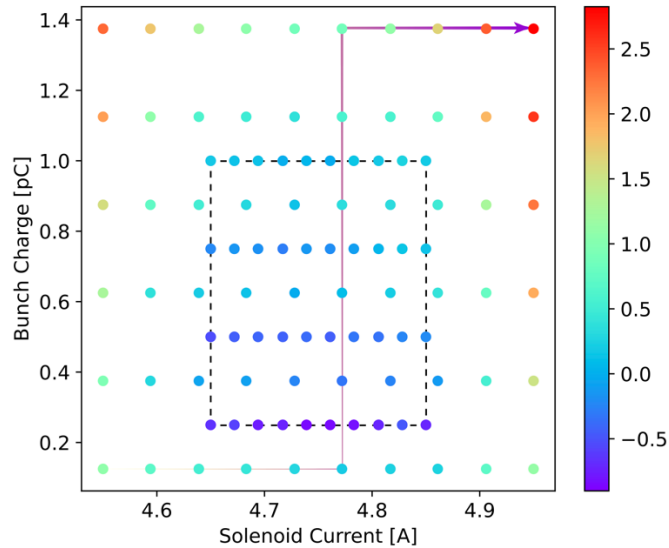

*Figure 7: The span of the training grid is shown within the dotted black square, with the closely spaced set of 4x10 grid points. The unseen extrapolation grid is the wider set of 6x10 grid points. The grid points are colored by percent mean absolute error on a log scale averaged over all points with **d** vectors at those locations. The arrow shows the path of interpolation movement.*

## *Adaptive cDVAE: Adaptive Latent Space Tuning for Time-Varying Systems*

We can take advantage of the low-dimensional latent embedding *c* which guides the diffusion process to develop an unsupervised adaptive application of the method which can track all 15 unique projections of a beam's 6D phase space based on just a single projection measurement as shown in Figure 9.

When cDVAE is applied in an unsupervised way, we assume that we have lost access to both *X* and *p*, which may have changed over time and the best we can do is replace them with our best guess based on previous data, which we refer to as $X_0$ and $\mathbf{p}_0$. The VAE's encoder then embeds $X_0$ into its low-dimensional latent space vector representation $z_0$. In many accelerators, while it is very difficult to accurately non-invasively measure $X_0$ and $p_0$ may also be unknown, or noisy, some downstream phase space measurements may be available. For example, in most high energy electron accelerators the (z,E) longitudinal phase space can be measured by using a combination of a transverse deflecting radiofrequency resonant cavity (TCAV) together with a dipole magnet and a scintillating screen. In free electron lasers (FEL) such as the LCLS, this measurement is considered to be non-invasive because it is performed after the electron beam has already generated the X-ray pulse that will be used for imaging and the light is not affected by the TCAV or the dipole which act on the electron beam. In other accelerators, sometimes there are scintillating screens which can be inserted to measure the (x,y) distribution as the beam passes through them and continues down the accelerator. If such a screen is far enough downstream to be at a high electron energy, then the distortion this measurement introduces in the beam's 6D phase space is minimal.

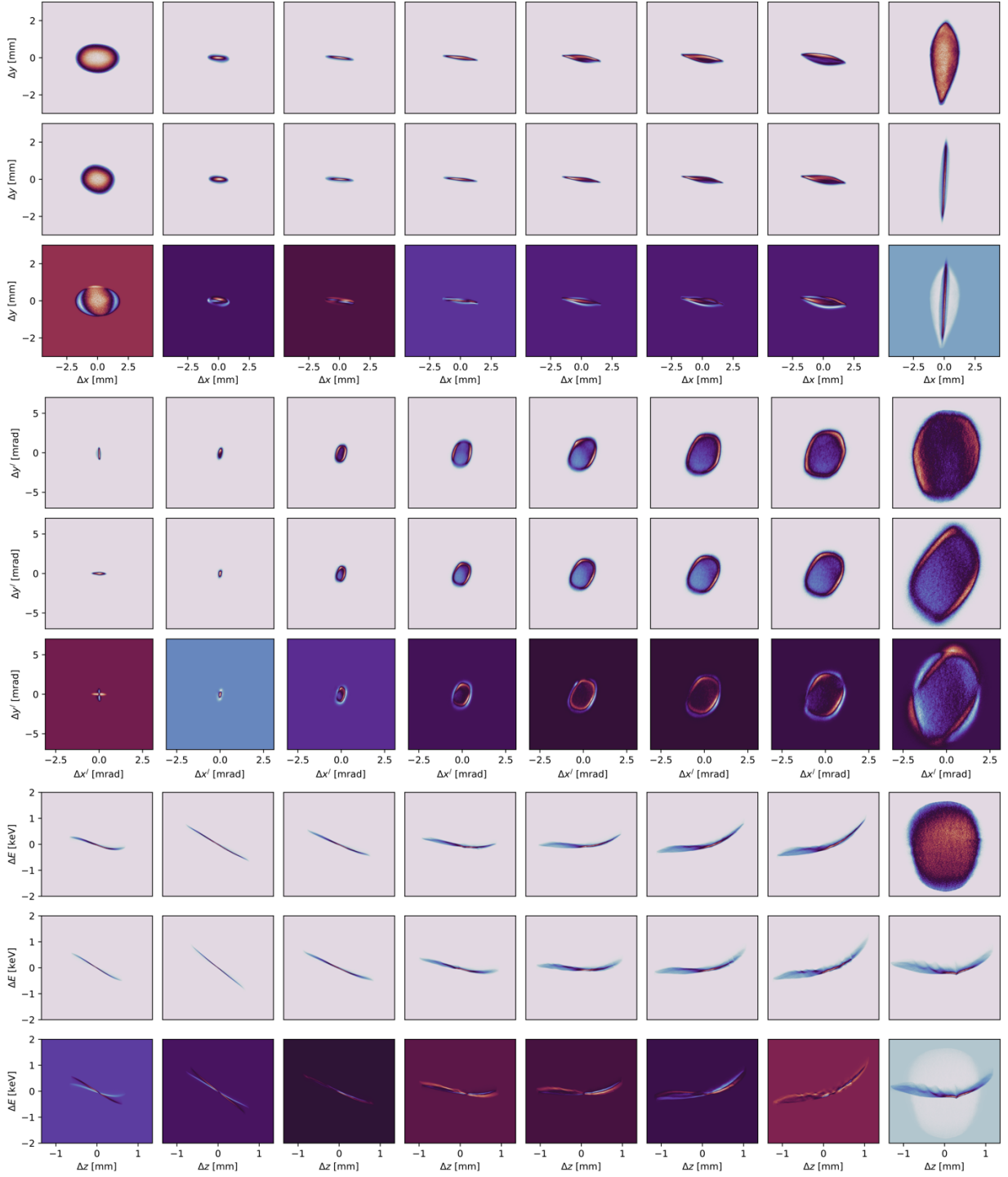

*Figure 8*: Each set of 3x8 subplots shows a comparison of predicted (top row) and true (middle row) phase space projections and their difference (bottom row) for the (x,y), (x',y'), and (z,E) projections at points where the path overlapped grid locations.

To illustrate the adaptive method, we consider the case where a non-invasive downstream measurement of a time-varying (z,E) distribution is available, referred to as *Y(t)*, is available, The adaptive cDVAE setup is shown in Figure 9. We create our initial conditional vector as $c_0=[z_0,p_0,d]$ where $d$=[000011] conditions the diffusion process to generate an estimate $\widehat{Y}(c_0)$ of the same distribution and compare it to the measured distribution to define a cost function

$$C(\mathbf{c_0}, t) = \int_z \int_E |Y(z, E, t) - \hat{Y}(z, E, \mathbf{c_0})| dE dz.$$

This cost function can be minimized to track the time-varying *Y(t)* by iteratively adjusting the value of *c*. The state-of-the-art method for minimization of analytically unknown, noisy, complex time-varying functions is a method known in the adaptive control community as extremum seeking (ES), which was originally developed for adaptive stabilization of unknown nonlinear dynamic systems by minimization of Lyapunov like functions [23, 39, 40]. To utilize ES in this case we leave *d*=[000011] fixed and adjust only the *z* and *p* components of *c* according to the ES algorithm:

$$\frac{dc_j(t)}{dt} = \sqrt{\alpha \omega_i} \cos(\omega_i t + kC(\mathbf{c}(t), t)),$$

where the $\omega_j$ are distinct perturbation frequencies, *k* is a feedback gain which can be used to speed up convergence, and $\alpha$ controls the dithering size. When utilizing the ES dynamics above, the average dynamics of the dithered components of *c* are given by

$$\frac{d\bar{c}_j(t)}{dt} = -\frac{k\alpha}{2} \nabla_{\bar{c}} C(\bar{\mathbf{c}}(t), t),$$

a robust gradient descent of the unknown function C (see [23], [39], or [40] for a proof). The relationship between the average trajectory and the actual trajectory of *c(t)* is that they can be made to stay arbitrarily close together for arbitrarily long periods of time for initial conditions over any compact set, by choosing sufficiently large values of the dithering frequencies $\omega_j$. Technically, for any compact set K, for any L>0 and any δ>0, there exists ω sufficiently large such that

$$\max_{t \in [0,L]} |\mathbf{c}(t) - \bar{\mathbf{c}}(t)| < \delta, \quad \forall \, \mathbf{c}(0), \bar{\mathbf{c}}(0) \in K, \quad \forall \omega_j > \omega.$$

Furthermore, if the averaged trajectory can be proven to track some trajectory *r*(t) in a robust and stable way for all *t*, then the above closeness of trajectories property can be extended to all time, resulting in the actual trajectory *c*(t) tracking *r*(t) arbitrarily closely (within an arbitrarily small δ) for all time [23, 39, 40].

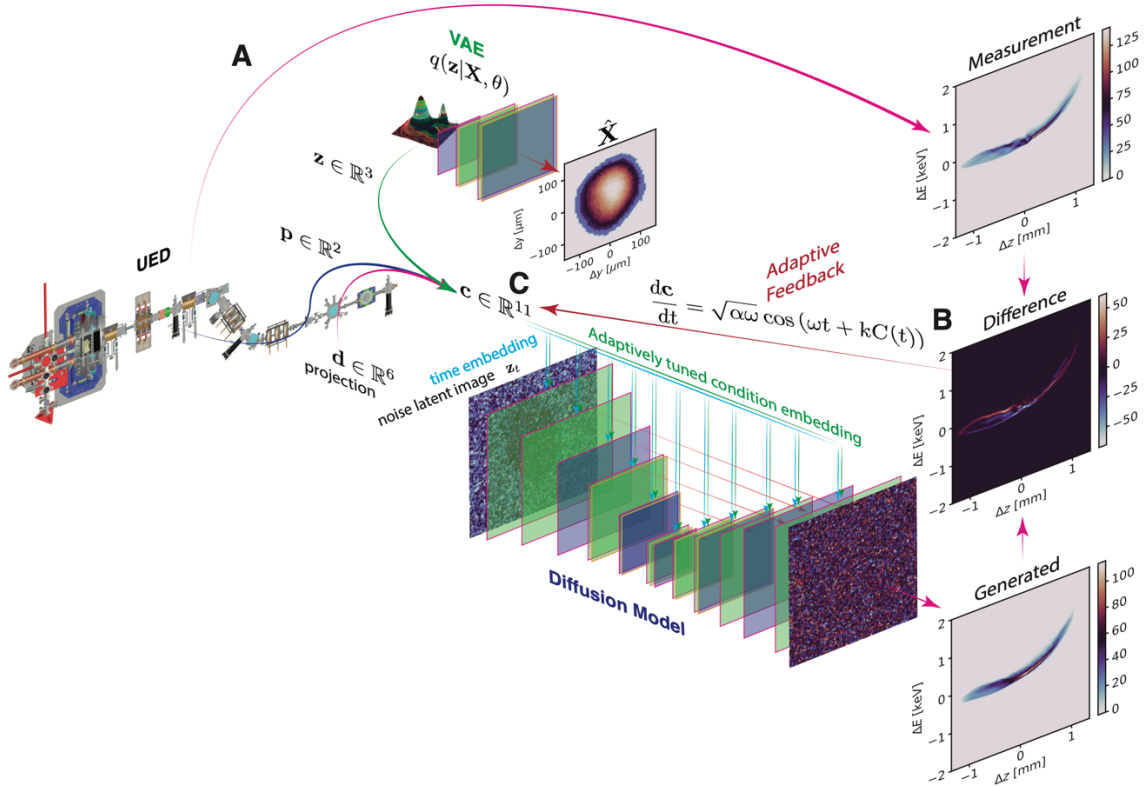

*Figure 9*: *Unsupervised adaptive application of cDVAE shown without access to measurements of initial beam conditions and with unknow scalar accelerator beam and parameter settings. We set those [z, p] inputs to our best guess and we set*

$d$=[000011] to generate a guess of the (z,E) distribution (A). The generated (z, E) is compared to its measurement (B). The difference guides adaptive feedback-based tuning of the conditional latent embedding (C).

Once these adaptive dynamics converge to and closely track a possibly time-varying minimum represented by latent point $c^*(t)=[z^*(t),p^*(t),d]$, which can be a local minimum, we then vary $d$ through all 15 possible choices, to generate all 15 projections of the beam's 6D phase space with the hope that by tracking (z,E), we are also tracking the other projections. In Figure 10, we demonstrate this approach by utilizing this ES-based tuning of the latent embedding to improve the predictions for the grid of data which includes points of extrapolation and compare them to the model's predictions without such feedback.

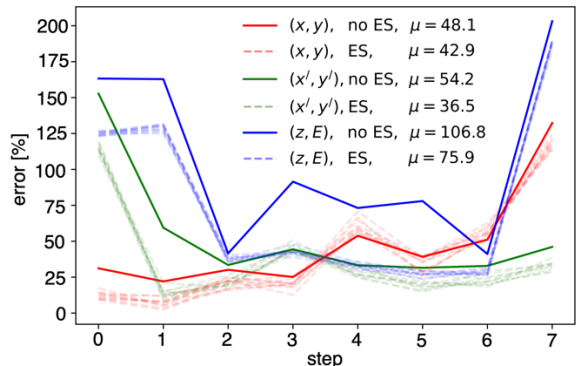

*Figure 10: Extrapolation and interpolation-based errors over the test set parameter grid was recorded while the network's conditional vector was iteratively adaptively adjusted via the (z,E) measurement-based ES method. Absolute prediction errors at various steps along the traversed grid of Figure 7 are shown. The solid lines show the best predictions for a known input. The dashed lines show the absolute prediction errors at the same locations, but after 60 steps of the ES algorithm were applied ten different times with ten different initial conditions.. It can be seen (blue) that the ES-based (z,E) prediction error is equal to or otherwise strictly less than the error based on a non-adaptive cDVAE prediction. This happens because the ES algorithm runs for 100 steps and then the location of the minimum cost value is used to record that latent embedding. The other adaptively tuned projections are sometimes better matches and sometimes worse than the original prediction based on $c_0$ because a choice of (z,E) that gives a local minimum of C does not necessarily provide the optimal (x,y) and (x',y') distributions since this answer is not unique.*

Figure 11 shows all 15 projections of the 6D phase space after adaptive feedback to improve the prediction accuracy for point 6 in Figure 10, resulting in a close match of generated projections to their true values.

## Conclusions

The cDVAE method has been shown to produce accurate representations of high-resolution images of all 15 projections of a charged particle beam's 6D phase space over a very wide range of conditions. This is exactly what generative diffusion-based models are state-of-the-art for; for generating very diverse highly complex high-resolution images. The cDVAE method utilizes multimodal (vectors + images) data can be combined into a single conditioning vector using a VAE. Furthermore, a physics-informed orthogonalization of the conditioning vector for each of the 15 projections has been developed to help the diffusion process learn the proximity of physically related projections. cDVAE is able to extrapolate beyond the training data, smoothly transitioning between various beam setups. This general method can be applied for any desired accelerator diagnostics and more generally for any generative problem for which data of various modalities need to be combined.

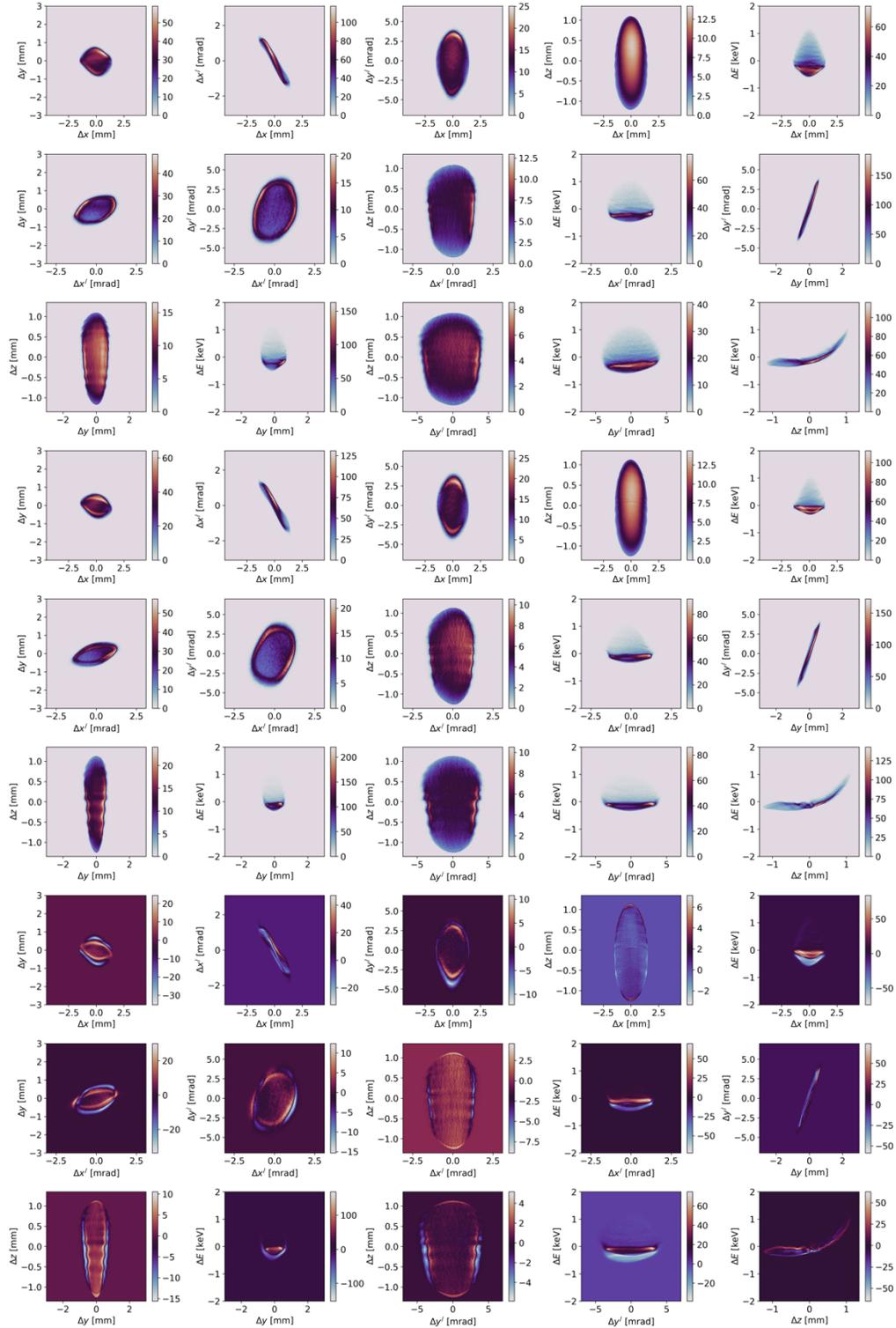

*Figure 11*: The top 3 by 5 grid of images shows the 15 adaptively tuned, generated phase space projections of the electron beam. The middle 3 by 5 grid shows their true values. The bottom 3 by 5 grid shows the difference. At this point cDVAE is extrapolating beyond the training data on a far edge of the grid, so as expected the created projections do not perfectly match their true values, with the best match being the bottom right corner (z,E) projection since that is the one projection that we comparing to a measurement in an attempt to track all other projections. Nevertheless, the overall prediction accuracy is high, mostly missing fine details, giving a good, highly accurate virtual view of the beam's phase space.

## Data Availability Statement

The data used for writing this paper is available from the lead author based on a reasonable request.

## Acknowledgements

This work was supported by the U.S. Department of Energy (DOE), Office of Science, Office of High Energy Physics contract number 89233218CNA000001 and the Los Alamos National Laboratory LDRD Program Directed Research (DR) project 20220074DR.

## Author contributions statement

A.S. conceived the experiment(s), A.S. conducted the experiment(s), A.S. analyzed the results. A.S. created and trained the neural networks, wrote and reviewed the manuscript.

## Competing Interests

The author declares no competing interests.